**On the flow and thermal characteristics of high Reynolds numbers (2800-17000) dye cell: simulation and experiment**

G. K. Mishra[1], Abhay Kumar[2], O. Prakash[1], R. Biswal, S. K. Dixit[1], S. V. Nakhe[1]

[1]Laser Systems Engineering Section,

[2]Proton Linac and Superconducting Cavities Division

Raja Ramanna Centre for Advanced Technology, Indore-452013

*E-mail: gkmishra@rrcat.gov.in

**Abstract**

This paper presents computational and experimental studies on wavelength/frequency fluctuation characteristics of high pulse repetition rate (PRR: 18 kHz) dye laser pumped by frequency doubled Nd:YAG laser (532 nm). The temperature gradient in the dye solution is found to be responsible for wavelength fluctuations of the dye laser at low flow rates ($2800 < R_{e^d} < 5600$). The turbulence Reynolds number ($Re_T$) and the range of eddy sizes present in the turbulent flow are found to be responsible for the fluctuations at high flow rates ($8400 < R_{e^d} < 17000$). A new dimensionless parameter, dimensionless eddy size ($l^+$), has been defined to correlate the range of eddy sizes with the experimentally observed wavelength fluctuations. It was found that fluctuations can be controlled by keeping $Re_T \approx 10$ and $l^+_{max} \approx 1$. The simulated result explains the experimental observation and provides a basis for designing the dye cells for high PRR pumping.

## 1. Introduction

High pulse repetition rate (PRR), tunable dye lasers have been utilized in atomic vapor laser isotope separation (AVLIS) schemes [1-3], holography [4] and medical applications [5-6]. A narrow and stable line-width of the dye laser together with a large average output power at high pulse repetition rate (PRR: 10s of kHz) improves the process efficiency of these applications [6-7]. These high repetition rate dye lasers are produced by interaction of a pump beam with a dynamic dye solution which circulates in a closed loop. The line-width is a strong function of wavelength fluctuations of the dye laser produced during the interaction of pump beam with the flowing dye solution [8-10]. The common high pulse repetition rate pump beam sources are copper vapor laser (CVL, 510 nm) [11], some advanced variants of CVL [12-14] and diode pumped solid state green lasers (frequency doubled Q-switched diode pumped Nd:YAG laser) [15]. The high repetition rate dye laser requires the dye solution to be flown at high speed through in the dye cell to remove the heat deposited by the pump beam [16]. In the recent past, significant efforts have been made to simulate the flow and thermal fields in a dye cell to develop a better understanding of the performance of high PRR dye laser [17-19]. A curved dye flow cell was analyzed for the dye solution flow in Reynolds number ($R_{ed}$)range of $184 < R_{ed} < 1100$ [17 − 18]. The authors also discussed the effects of eddy sizes on dye laser bandwidth. In another study, a straight dye flow cell was analyzed in $1000 < R_{ed} < 7000$ flow regime for obtaining spatial trends of temperature and kinetic energy of fluctuating part of velocity [19]. These analyses provided explanations for the observed trend of bandwidth and wavelength variation of 5.6 kHz PRR CVL pumped dye laser.

This paper presents numerical simulation using computational fluid dynamics (CFD) software ANSYS Fluent release 15.0 [20] to compute thermal and flow fields in turbulent flow regime ($2800 < R_{ed} < 17000$) of dye solution (1 mMRh6G dye in ethanol). The paper also presents a new approach based on turbulence Reynolds number ($Re_T$) [21] and eddy size present in the turbulent flow to explain the wavelength stability of the dye lasers. The work will be very useful in optimizing the geometry and dye solution flow rates for improved performance in terms of a narrower line-width of dye lasers operating at high PRR based on $Re_T$. The simulated results have been tested in a study on wavelength fluctuations of a 18 kHz PRR frequency doubled (532 nm) Q-switched Nd:YAG laser pumped dye laser against dye solution flow rate. A good agreement between simulations and experimental results has been observed.

## 2. Flow considerations

A forced convection flow field is characterized by Reynolds number [21] given in equation (1):

$$R_{ed} = \frac{u d_h}{\nu} \quad (1)$$

Where $u$ and $d_h$ are the average flow velocity and hydraulic diameter of the flow channel respectively. The hydraulic diameter is given in equation (2):

$$d_h = \frac{4A}{P} \quad (2)$$

Where $A$ and $P$ are cross sectional area and wetted perimeter of the flow channel respectively. A flow regime of $R_{ed} > 2300$ is turbulent that offers high heat transfer rates [22]. About 20 to 25% of optical pump beam power is converted into heat in the dye solution by the processes of (i) radiationless deactivation of dye molecules in excited states and (ii) Stoke shift [23-24]. The heat deposited in the dye solution by the pump laser beam follows an exponential law of equation (3) [25]

$$Q^* = Q_o^* e^{-\frac{y}{y_o}} \quad (3)$$

Where $Q^*$ is the volumetric heat deposition at a distance $y$ from the wall, $Q_o^*$ is the volumetric heat deposition at the wall and $y_o$ is the penetration depth at which the heat density reduces to 1/e times of that at the wall. When $Q_o^*$ is multiplied by the volume of gain region up to the penetration depth (~100 μm); it gives the total amount of heat absorbed in the gain medium.

It can be seen from equation (3) that about two-third of the deposited heat lies within a penetration depth of about 100 μm from the wall of the pump beam entrance window of the dye cell. This region is known as gain medium region and interactions within this region are mainly responsible for the optical properties of the dye laser. The variation of dye laser wavelength depends on the refractive index of material [26] and is given in equation (4):

$$\frac{\Delta \lambda}{\lambda} \approx \frac{\Delta n}{n} \qquad (4)$$

Where $n$ is the refractive index of the dye solution and $\Delta n$ is its change.

The term $\frac{\Delta n}{n}$ depends on the temperature uniformity as well as on the size of eddies in the gain medium region. The temperature of the dye solution in this region is governed by exponentially decaying heat deposition by the pump beam and the heat transport by the forced convection. Therefore, this will set up a temperature gradient in the dye solution normal to the wall as well as in the direction of flow. The refractive index of the dye solution is a function of temperature [16, 26]. The wavelength fluctuation of the dye laser depends on the temperature gradient of the solution in the gain medium region [26] and is given in equation (5) and (6):

$$\frac{1}{\lambda}\frac{d\lambda}{dT} = \frac{1}{n}\frac{dn}{dT} \qquad (5)$$

$$\Rightarrow \left(\frac{\Delta \lambda}{\lambda}\right)_{thermal} = \frac{1}{n}\frac{dn}{dT}\Delta T \qquad (6)$$

The temperature variation in the gain medium region predominantly depends on the flow velocity of the dye solution [16, 24]. The flow becomes highly turbulent at higher flow velocities and leads to a decrease in the temperature variation in the gain medium region due to increased turbulent mixing. However, these highly turbulent flows contain eddies of different sizes. Each eddy of a particular size of a turbulent flow can be considered to be homogeneous in refractive index [27]. The continuous distribution of eddy sizes between the large scale length and the small scale length leads to a distribution of refractive indices following an inverse power law of the physical size of the eddies [27].

The refractive indices of eddies are functions of eddy size and follow the relation given in equation (7):

$$n \propto \frac{1}{r^m} \tag{7}$$

Where *m* is a constant and *r* is the size of the eddy.

This leads to a simple statement of equation (8):

$$\frac{\Delta n}{n} \propto \frac{\Delta r}{\bar{r}} \tag{8}$$

Where $\Delta r$ is the difference between the largest eddy size (*l*) and the smallest eddy size ($\eta$) and $\bar{r}$ is the average eddy size. We have designated their ratio as dimensionless eddy size ($l^+$) which is given in equation (9):

$$l^+ = \frac{\Delta r}{\bar{r}} \tag{9}$$

The wavelength fluctuation in presence of turbulence can be estimated from equations (8) & (9)

$$\left(\frac{\Delta \lambda}{\lambda}\right)_{turbulence} \propto l^+ \tag{10}$$

The large eddies (*l*) are unstable and break-up, transferring the fluctuating part of the kinetic energy, popularly known as turbulence kinetic energy (*k*) to somewhat smaller eddies. These smaller eddies undergo a similar break-up process and transfer their energy to yet smaller eddies. This energy cascading process in which energy is transferred to successively smaller and smaller eddies; continues till the viscosity is effective in dissipating the turbulence kinetic energy. At these smallest eddy sizes ($\eta$), the turbulence kinetic energy ($k$) is converted into heat. Reynolds numbers are the ratios of inertia forces and viscous forces. Therefore, eddies won't survive when turbulence Reynolds number ($Re_T$) [20], based on the characteristic length of large eddy size (*l*), approaches unity *i.e.* the eddy sizes have physical meaning only when this Reynolds number is more than unity. $Re_T<1$ is the region where eddies are absent. At very large turbulence Reynolds numbers ($Re_T>>1$), there will be a significant difference between the largest eddy size and the smallest eddy size. The largest eddy size (*l*), turbulence Reynolds number and smallest eddy size are given in equations (11), (12) and (13) respectively [20, 28, 29]

$$l \cong \frac{k^{\frac{3}{2}}}{\epsilon} \tag{11}$$

$$Re_T = \frac{u'l}{\nu} = \frac{l}{\nu}k^{\frac{1}{2}} \tag{12}$$

$$\eta = \frac{l}{Re_T^{0.75}} \tag{13}$$

Where, $\epsilon$ is the rate of dissipation of the turbulence kinetic energy, $u'$ is the root mean square of the fluctuating part of the velocity and $\nu$ is the kinematic viscosity of the fluid. The largest eddy size depends on the geometrical extent of the flow channel [20]; therefore the dimensions of the gain medium region are very important to determine the largest eddy size, turbulence Reynolds number and smallest eddy size.

The turbulence Reynolds number varies throughout the flow field and would increase as we move away from the wall due to a decrease in the viscous forces. Eddies would not be found in near-wall region where essentially a laminar sub-layer exists [30]. As we increase the flow, the thickness of the laminar sub-layer decreases and higher turbulence Reynolds number would appear in the near-wall region. The extent of variation of dye laser wavelength, as suggested by equation (4), will be determined by equations (6) and (10) and can be represented in equation (14) as:

$$\left(\frac{\Delta\lambda}{\lambda}\right)_{total} = \left(\frac{\Delta\lambda}{\lambda}\right)_{thermal} + \left(\frac{\Delta\lambda}{\lambda}\right)_{turbulence} \tag{14}$$

## 3. CFD model of dye cell

Fig 1 shows the schematic of dye cell. The flow cross-section of the flat pinched region was 0.7 mm x 15 mm. The transverse pumping by the pump beam was through a window of 0.5 mm x 15 mm at the center of the dye cell. The near-wall region is important for investigation as the maximum absorption of the pump beam takes place near the wall. Wilcox $k$–$\omega$ shear stress transport (SST) turbulence model [31-32] is highly appropriate to calculate the flow and thermal field near the wall without losing the accuracy of the flow behavior away from the wall. This hybrid model combines the Wilcox $k$–$\omega$ model near the wall and the $k$–$\varepsilon$ model away from the wall. Two blending functions, F1 and F2, calculated by the software on the basis of the distance from the nearest surface and the flow variables, ensure that the appropriate model is utilized throughout the flow field [20].

ANSYS Fluent supports the simulation of two-dimensional and three-dimensional flow fields. A flow field can be considered to be two-dimensional if the effect of walls in the depth direction can be neglected i.e. d >> t (fig.1). The hydraulic diameter for a truly 2D case will reduce to equation (15):

$$d_{h-2D} = \frac{4\,(d \times t)}{2(d+t)} \approx \frac{4\,(d \times t)}{2(d)} = 2t \tag{15}$$

For the dimensions of the dye cell under investigation, $d_h$ is 1.34 mm and $d_{h-2D}$ is 1.4 mm. This indicates that the CFD model can be solved as a 2D case. Both 3D as well as 2D flow field was solved in order to verify this assumption. 2D simulation gives flexibility to design the mesh topology to accurately capture the temperature and eddy sizes in the near-wall region. The finite volume mesh near the wall surface has been kept very fine to obtain dimensionless wall distance [20, 25] of 0.5 and near-wall element size of 6 μm. The computer hardware used in the simulation was HP Z800 workstation with 2p-12c Xeon X5660 processors, 48 GB RAM and 1 NVIDIA Tesla C2075 GPU. ANSYS Fluent's parallel processing option was used for solution.

    The total pump beam power has been considered to be 9 W. The total deposited heat in the fluid volume of the gain medium region is estimated to be 2 W which is about 22.5% of the total beam power. This exponentially decaying volumetric heat source was applied in the dye gain region of 0.5 mm × 0.7 mm using a user defined function(UDF) written in the C programming language. This function is dynamically loaded with the ANSYS Fluent solver during the computation. The Reynolds number given by equation (1) is kept above 2300 [21] to ensure a turbulent flow which facilitates a good heat transfer and a lower temperature variation in the dye solution. The analyses were performed at flow rates of 2, 4, 6, 8, 10 and 12 LPM. Fig. 2 gives the calculated values of Reynolds numbers for these flow rates. It is seen that the Reynolds number varied from 2800 – 17000 for the flow rates from 2 LPM to 12 LPM. The diverging part of the dye cell was initially modeled to see if the effects of vortex formations in the diverging part are travelling to the gain medium region. After ascertaining that these effects do not travel to the pinched region, we truncated the CFD model to keep the convergent portion and flat pinched region portion to run the analysis faster. The modeling of converging part is important to allow the flow to develop completely before it reaches the pinched region.

## 4. Computed results

Fig. 2 shows the Reynolds number in the gain medium region for flow rates varying from 2 LPM to 12 LPM. An investigation of the velocity profile along the flow direction in the pinched region shows that the flow is fully developed for all flow rates from 2 LPM to 12 LPM. Fig. 3shows the fully developed velocity profiles in the pinched region at the flow rates of 2, 4, 8, 10, 12 LPM.

The temperature of the gain medium rises in the vicinity of wall due to large deposition of heat in the near wall region. Fig.4 shows that temperature contours for 2 LPM, 6 LPM and 12 LPM flow rates. One can see that the temperature rise also takes place in the flow direction as more and more heat gets deposited in the medium as it progresses. However, the temperature gain falls in the direction normal to the wall due to exponential decay in the heat deposition. Fig. 5 shows temperature rise in the direction normal to the wall at the exit of gain medium region. The maximum temperature rise ($\Delta T_{max}$) of ~ 2.1 K takes place when the flow is 2LPM and decreases monotonically to ~0.15 K at 12 LPM.

Fig. 6 shows the spatial variation of turbulent kinetic energy along the pump beam direction in the dye cell for the dye solution flow rates from 2 to 12 LPM. It is seen that as the flow is increased, the turbulence inside the near-wall region of the gain medium also increases. The maximum of turbulent kinetic energy shifts towards the wall as the flow rate is increased and rises sharply for flow rate >6 LPM. This indicates that the effect of turbulence may dominate for the flow rate > 6 LPM. The maximum values of the turbulent kinetic energy varied from 0.03 J/kg to 2.5 J/kg for the flow rate of 2 to 12 LPM. The turbulence produces eddies of various sizes. The generated largest eddies size (*l*) for various dye solution flow rates (2 to 12 LPM) is shown in fig.7. It is clear that the magnitude of $l$ in the gain region (~ 100 µm) varied from 10 µm to 50 µm as the flow rate is changed from 2 to 12 LPM.

Fig. 8 shows the thicknesses of laminar sub-layer and buffer layer [25] together with a layer where $Re_T$ just exceeds unity for the flow rate of 2 LPM to 12 LPM. It is seen that the $Re_T$ layer is very close to the laminar sub-layer. This is in line with the predictions of boundary layer theory which says that effects of turbulence are experienced in the buffer layer [30]. This indicates that eddies are present in between laminar sub layer and buffer layer.

Fig.9 shows the variation of $Re_T$ against distance from the wall in the gain medium region. For flow rate of 2 LPM ($R_{ed} \approx 2800$), $Re_T$ is less than unity, therefore stable eddies will not be formed. At 4 LPM it approaches unity above the distance of ~ 45 µm, from the wall, which covers only half of the gain medium region (~ 100 µm). For the flow rates above 4 LPM ($R_{ed}$>5600) $Re_T > 1$ in some parts of the gain medium region, therefore $l^+$ exists but its value exceeds unity only after 6 LPM flow rate. The variation of $l^+$ in the gain medium region is shown in fig. 10. The maximum values of $l^+$ ($l^+_{max}$) for the flow rates of 4, 6, 8, 10 & 12 LPM are around 0.5, 1.5, 1.7, 1.77 & 1.82.

The variations of dye laser performance characterizing parameters i.e. $\Delta T_{max}$ and $l^+_{max}$, with the Reynolds numbers ($R_{ed}$) from 2800 to 17000 corresponding to the flow rates of 2 to 12 LPM is shown in fig.11. It is seen that the temperature gradient is dominant for lower flow rate (< 4LPM, $R_{ed}$~5600), whereas the eddy size is dominant at higher flow rates. Another crucial observation can be made that the location of $l^+_{max}$ from the wall also remains nearly constant of 60 µm at 10 LPM and 12 LPM flow rates. This indicates that the effect of turbulence is nearly same for the flow rates between 10-12 LPM.

## 5. Experimental study

Fig. 12 shows schematic of experimental set up of ~18 kHz dye laser. The dye laser optical cavity was a standard grazing incidence grating (GIG) resonator of the length ~16 cm. The cavity consisted of a grating of groove spacing ~2400 l/mm, a 4% output coupler wedge, a fully reflecting mirror (R > 99%) and a double prism beam expander (M~ 22 at angle of incidence ~80°). Dye solution of 1mM concentration of laser grade Rh-6G dye in the ethanol was circulated through a variable flow circulator system. Dye solution flow rate was varied from 2 to11 LPM. A homemade frequency doubled diode pumped Q-switched Nd:YAG solid state laser [15]was used to transversely pump the dye laser with help of a cylindrical lens of focal length 50 mm. The pump laser ($\lambda$= 532 nm) was operated at 18 kHz pulse repetition rate at average output power of ~9 W. The dye solution reservoir temperature was maintained at 22± 0.5°C by a water cooled heat exchanger coil submerged in the dye solution.

The wavelength of dye laser and its fluctuations were recorded for different flow rates (2 to 11 LPM). Fig.13shows the snapshot of the recorded wavelength fluctuations from a wave-meter (High Finesse: WS-7) at different flow rates. For each flow rate the data was recorded for about 10 minutes. The measured values of these wavelength fluctuations with different flow rates is shown in fig. 14.A large fluctuation of ~0.005 nm was observed at 2 LPM($R_{ed} \approx 2800$) which reduced to ~ 0.0025 nm in the flow range of 4 to 6 LPM (5600<$R_{ed}$ < 8400) and it again increased to ~0.004 nm as the flow rate increased from 6 to11 LPM(8400<$R_{ed}$ < 15500).

## 6. Discussion

The above experimental results can be explained on the basis of two flow related parameters - temperature gradient ($\Delta T$) and dimensionless eddy size in the gain medium region. For simplicity of discussion, the investigated flow regime can be divided into three different regions i.e. region-1: from 2 to 4 LPM, region-2: from 4 to 6 LPM and region-3: from 6 to 11 LPM (fig. 14). For the flow rate of 2 LPM ($Re_T < 1$), the wavelength fluctuations can be explained on the basis of a significant temperature rise in the gain medium region as eddies are not developed. This is evident from the fig. 11, where the corresponding estimated temperature rise of the dye solution is ~ 2.1 K and $l^+$ does not exist in the gain medium region. The temperature-maximum occurs at the wall where the maximum energy is also deposited by the pump beam absorption. Since the energy deposition by the pump beam is responsible for the temperature rise of the dye laser gain medium, the relative significance of proximity of this temperature rise near the wall is very high. Therefore this flow regime can be termed as region of thermal field instability. At 4 LPM flow rate ($R_{ed} \approx 5600$); the $Re_T > 1$ after ~ 45µm but its maximum value within the gain medium region is ~ 2. The effect of viscous forces is still very high and $l^+$ values are very small (0.53 at 90µm). The temperature rise in the gain medium region has decreased to ~ 1.3K. Therefore a better stability is expected and is seen in the experiments.

At 6 LPM ($R_{ed} \approx 8400$), the CFD computations show that although the temperature rise has reduced to 0.6K but the turbulence Reynolds number ($Re_T$) is now larger than ten for a large part of the gain medium region (from 45 to 95 µm from the wall). Values of $l^+$ are also larger than unity i.e. the difference in the largest and the smallest eddy sizes is larger than the average eddy size beyond 26 µm from the wall. The experiments also showed that the wavelength fluctuations increase at this flow rate. Therefore, the wavelength fluctuations should be lowest in the flow regime of 4 LPM to 6 LPM ($5600 < R_{ed} < 8400$). This can be seen in the experimental results. At 8 LPM flow rate ($R_{ed} \approx 11200$), the temperature rise has reduced to 0.35K but $Re_T > 10$ for a large part of the gain medium region (from 23 µm onwards) and its maximum value reaches 30 at ~ 60 µm from the wall. $l^+$ values also increase and remain above unity beyond a distance of 15 μm from the wall. $l^+_{max}$ is 1.71 at 65 μm from the wall. Therefore, a significant part of the gain medium region remains under the influence of eddies and get affected by the variation in their sizes. We can see in the experiments that the wavelength fluctuations increase at 8 LPM. Beyond 8LPM, the above mentioned trend continues but with a small difference. Both $Re_T$ and $l^+$ increase in the gain medium region but their rate of rises are different. The increase in $l^+_{max}$ is very small and proximity of $l^+_{max}$ with the wall also approaches ~60 μm from the wall. One would expect the effect of turbulence to remain nearly constant at flow rates between 10-12 LPM ($14000 < R_{ed} < 17000$). This explains the nearly same wavelength fluctuation for 10 LPM and 12 LPM flow rates observed during the experiments (fig.14). Hence region-3 can be attributed as region of flow field instabilities, whereas region 2 can be attributed as a stable performance region.

An important design guideline can be generated on the bases of the CFD computations and the experimental observations. In order to bring down the thermal inhomogeneity, the flow rate of the dye solution should be increased till $Re_T \approx 10$ in the gain medium region. At this value of $Re_T$, the $l^+$ values should be checked and $l^+_{max}$ should be kept below unity in the gain medium region by minor modification in the flow rate. The final value of flow rate can be determined experimentally by tuning the same around the calculated value.

## 7. Conclusion

CFD computations were performed to analyze the flow of the dye solution. The temperature gradient in the dye solution were computed and found to be responsible for wavelength fluctuation of the dye laser at low flow rates ($2800 < R_{ed} < 5600$). The temperature gradient diminishes at higher flow rate but the turbulence of the flow starts affecting the wavelength fluctuations. At high flow rates ($8400 < R_{ed} < 17000$), two dimensionless parameters of a turbulent flow, turbulence Reynolds number($Re_T$) and dimensionless eddy size ($l^+$) are used to correlate the computed results with the experimental findings. It was found that the wavelength fluctuations can be controlled by keeping $Re_T \approx 10$ and $l^+_{max} \approx 1$ in the gain medium region.

**Acknowledgements**

Authors sincerely acknowledge Dr. S. M. Oak, Head SSLD, RRCAT for providing motivation and his keen interest in the work. The authors wish to acknowledge the help of Mr. S. C. Joshi for supporting the work at Proton Linac and Superconducting Division. Participation of Mr. A. J. Singh and Mr. S. K. Sharma in the experiments is gratefully acknowledged. Authors would like to acknowledge the contributions of Mr. S. K. Agrawal, Mr. U. Kumbhkar and Mr. D. Joshi in setting up dye laser experimental facility.

**List of figure captions**

**Figure-1:** Schematic of Dye cell

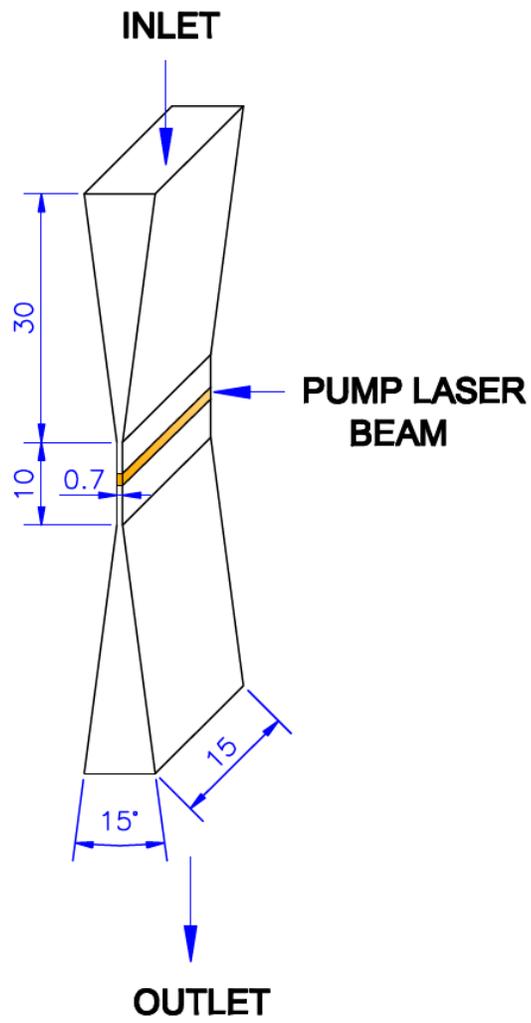

**Figure-1**

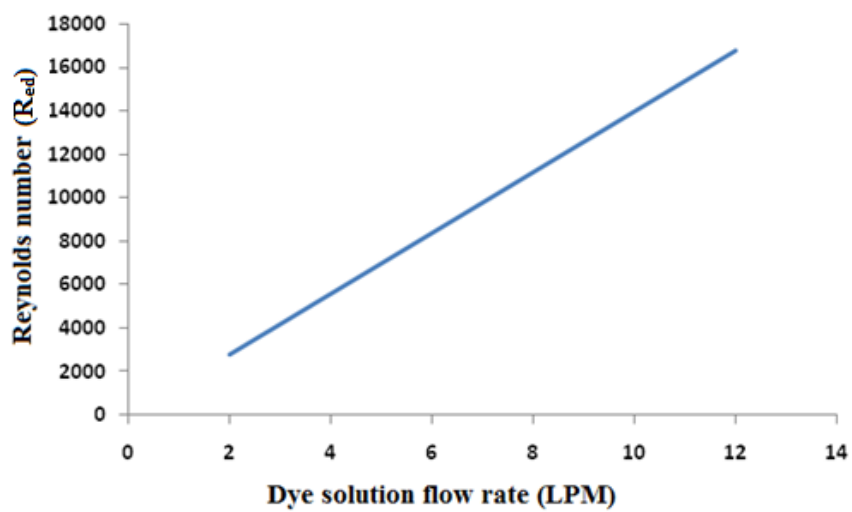

**Figure-2**

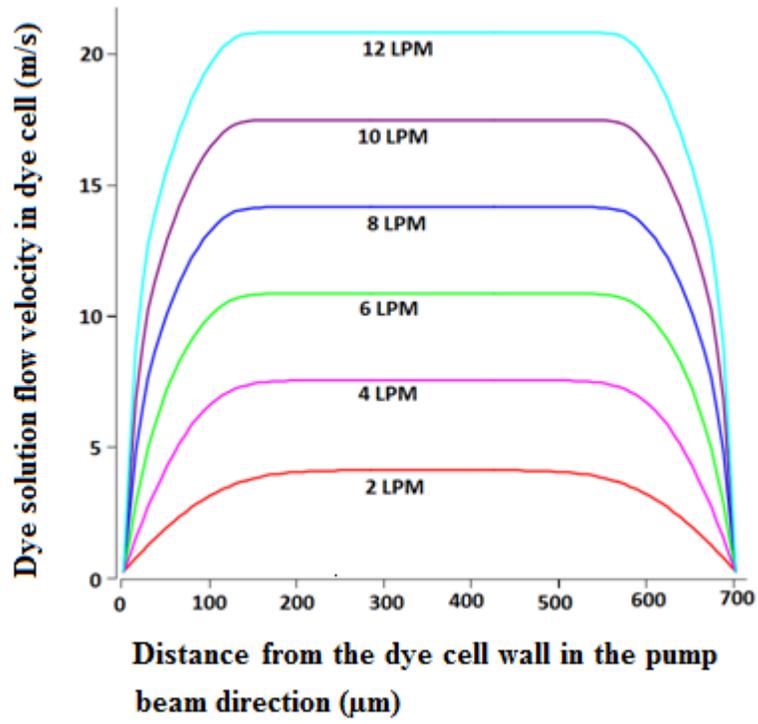

Figure-3

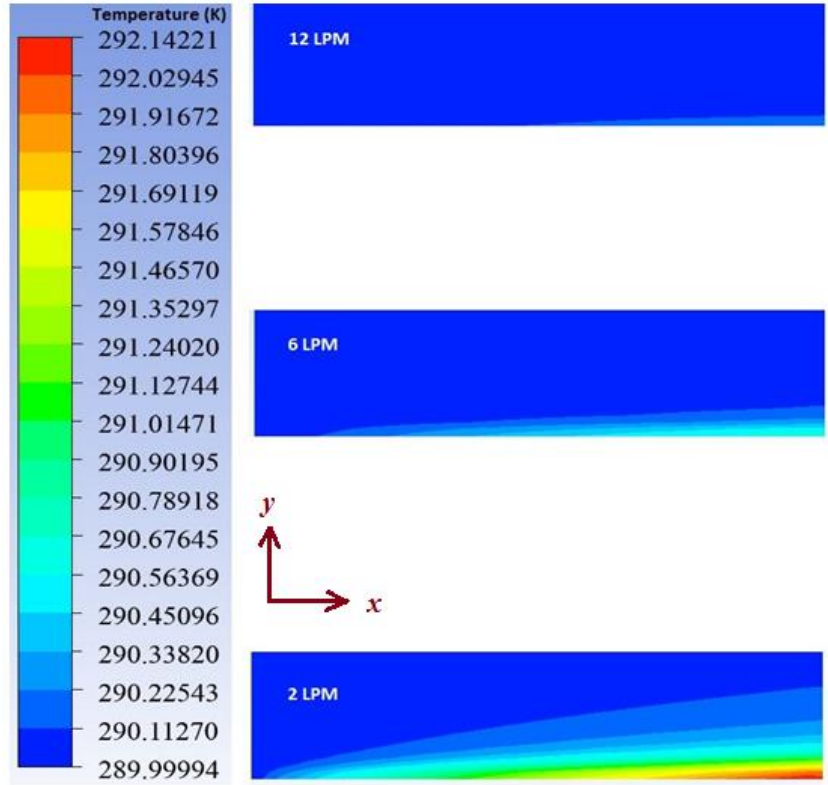

**Figure- 4 Temperature contour of dye gain medium**

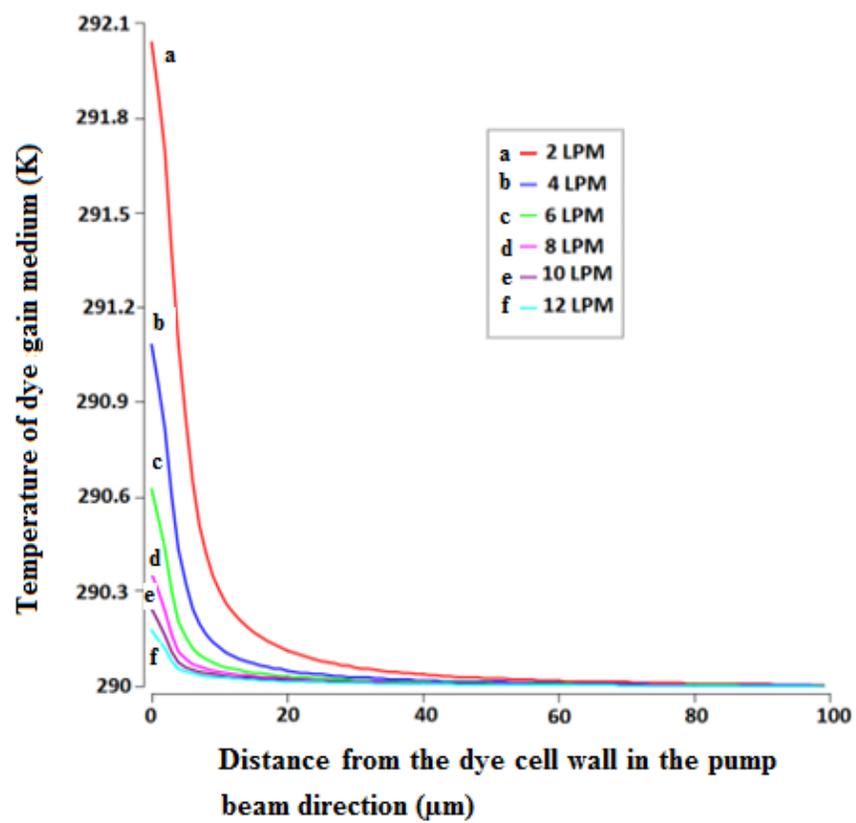

**Figure-5**

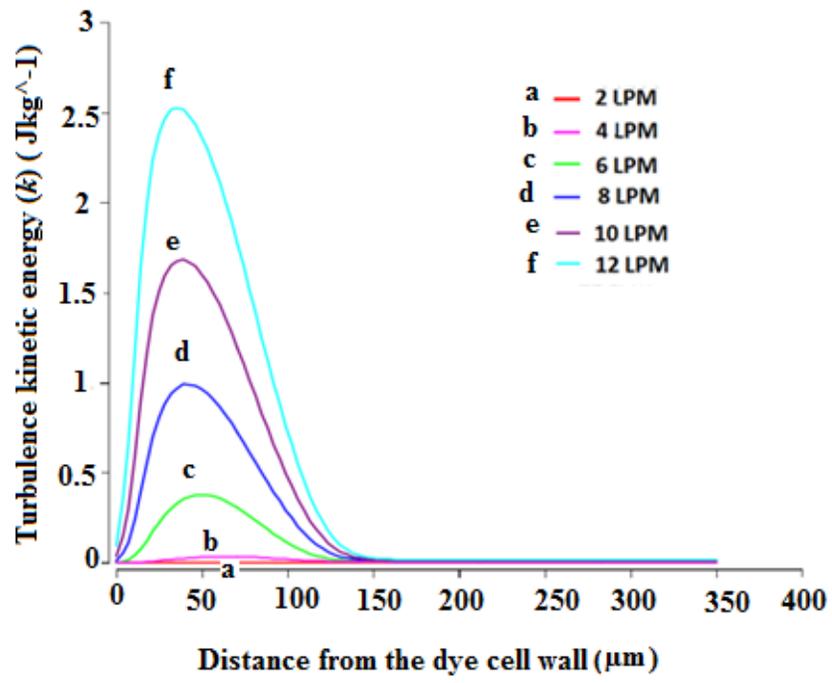

**Figure - 6**

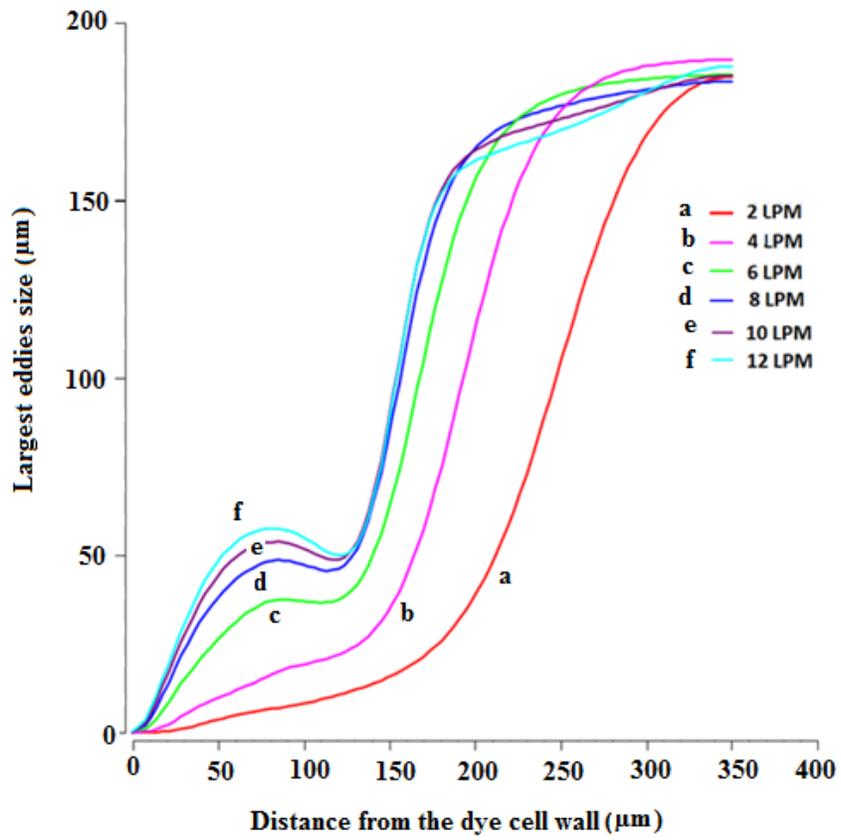

**Figure -7**

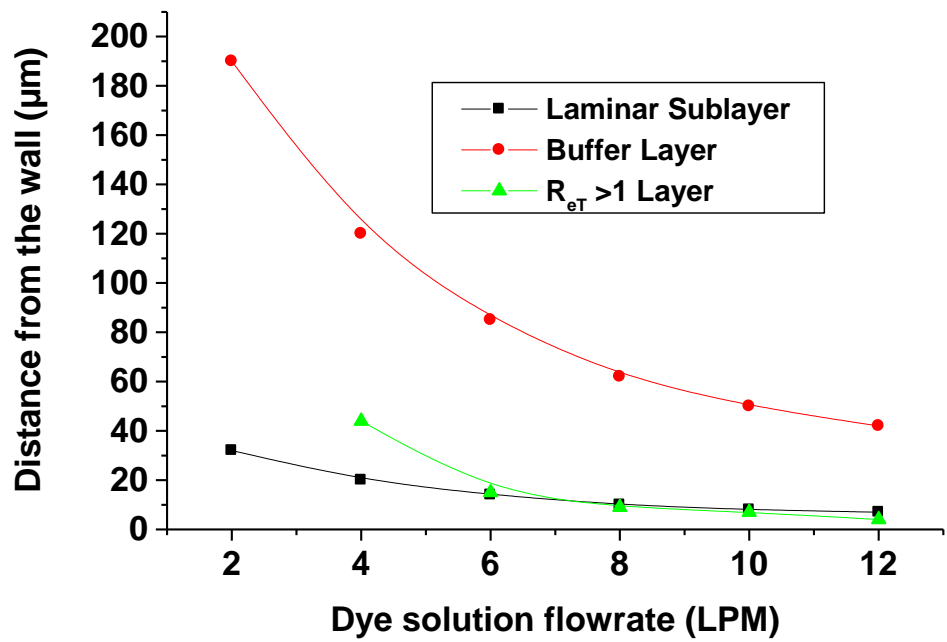

Figure -8

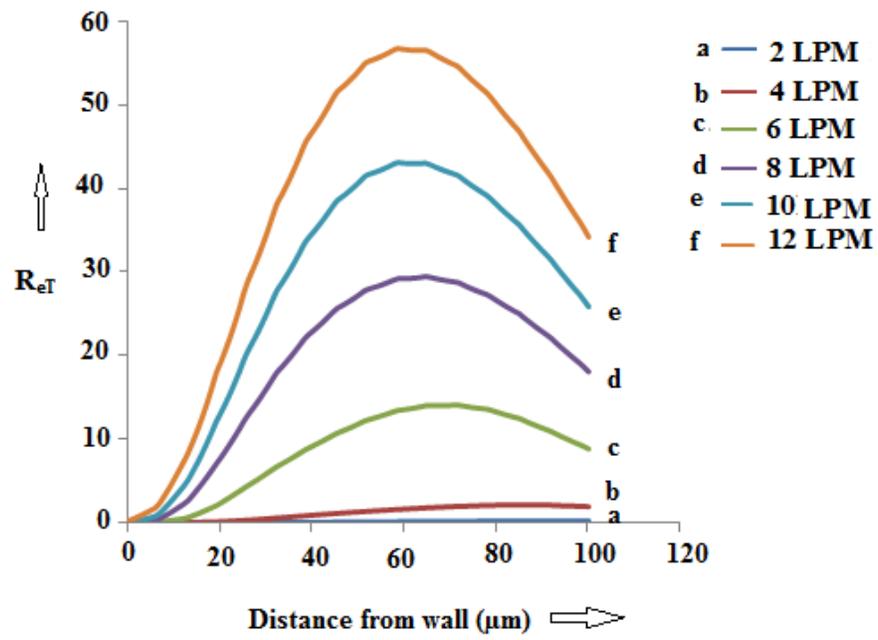

**Figure -9**

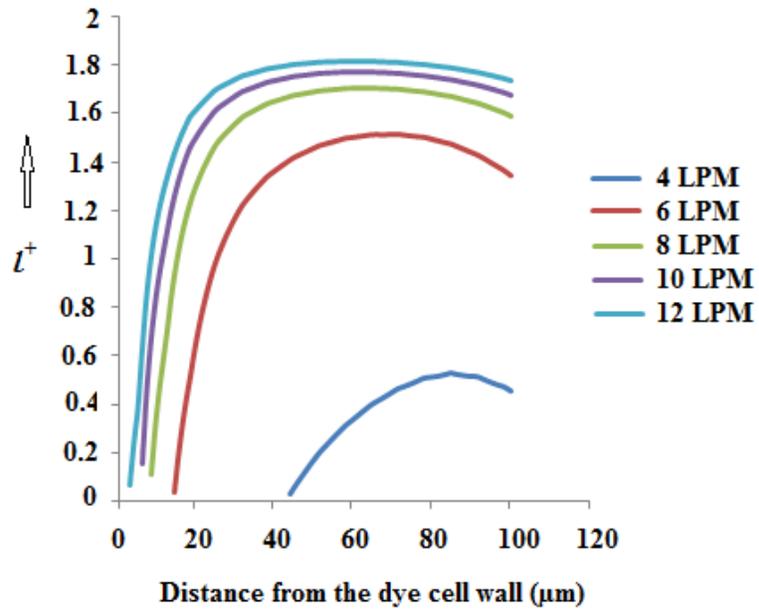

**Figure -10**

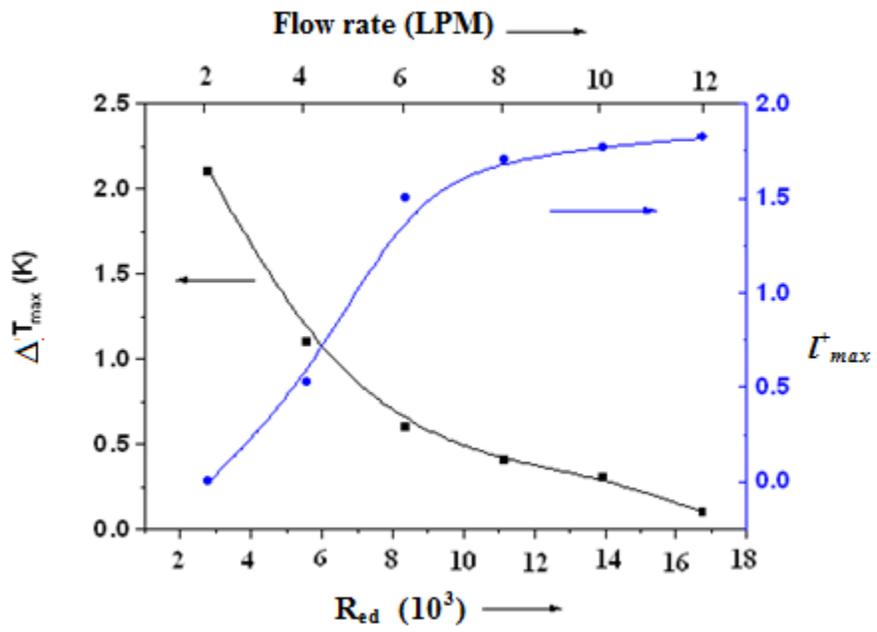

**Figure-11**

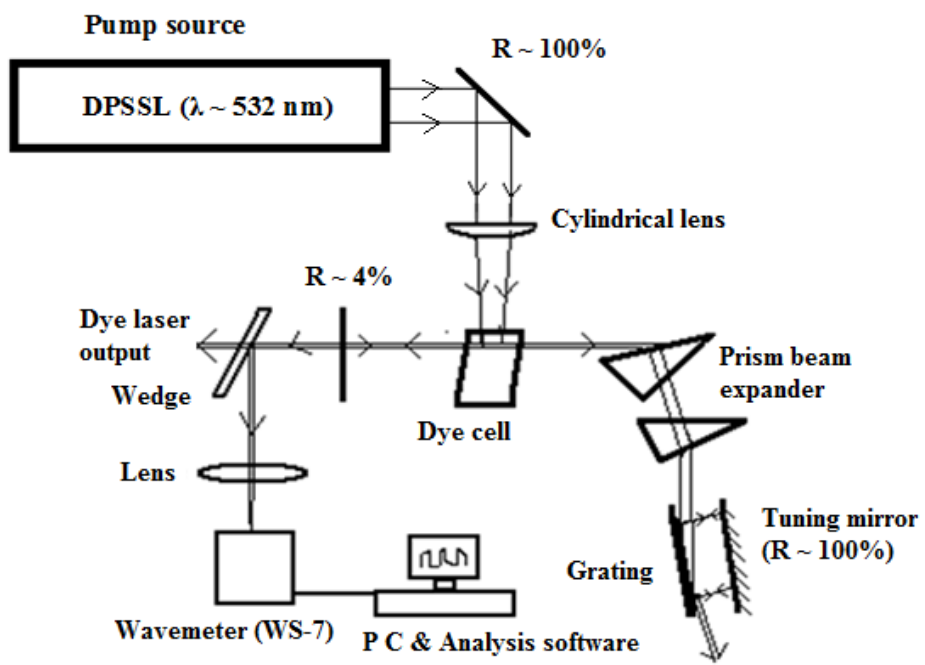

**Figure - 12**

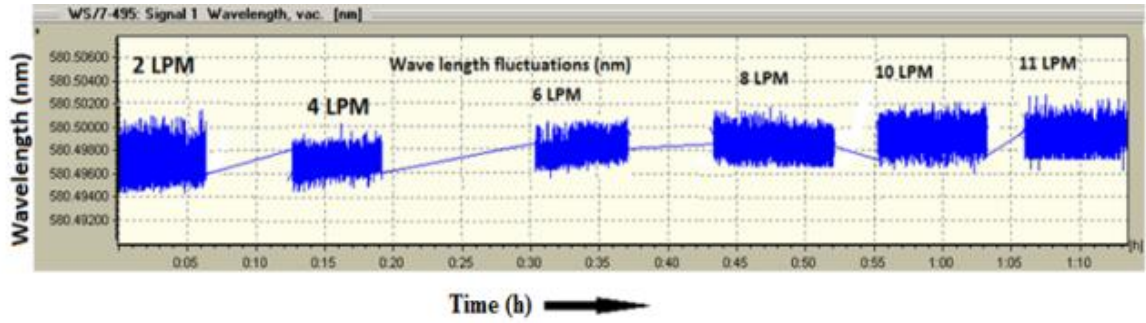

**Figure – 13**

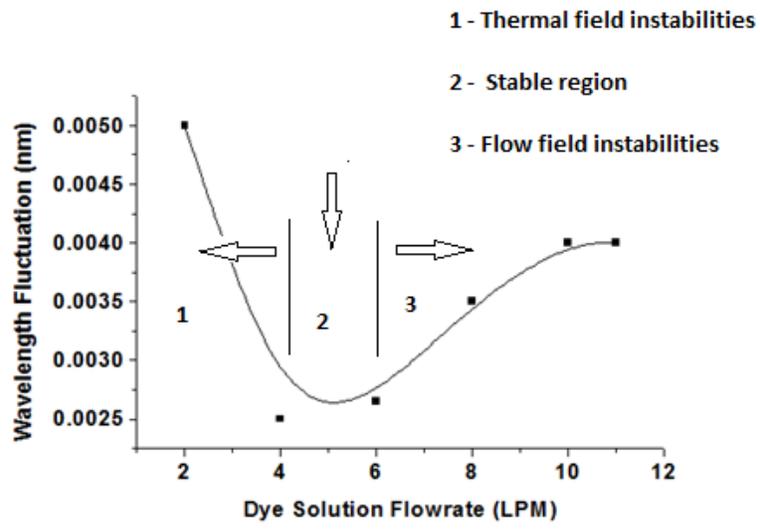

**Figure - 14**